\documentstyle[prl,aps,twocolumn,floats,epsf]{revtex}
\tighten
\begin{document}

\title{Dibaryons with Strangeness: their Weak Nonleptonic Decay using SU(3)
  Symmetry and how to find them in Relativistic Heavy-Ion Collisions}

\author{J\"urgen Schaffner-Bielich}
\address{RIKEN BNL Research Center,
Brookhaven National Laboratory, 
Upton, New York 11973-5000, USA}

\author{Raffaele Mattiello}
\address{The Niels Bohr Institute, 
Blegdamsvej 17, 
DK-2100 Copenhagen, Denmark}

\author{Heinz Sorge}
\address{Department of Physics \& Astronomy, State University of New York at
  Stony Brook, NY 11794-3800, USA}

\date{\today}
\maketitle

\begin{abstract}
Weak SU(3) symmetry is successfully applied
to the weak hadronic decay amplitudes of octet hyperons.
Weak nonmesonic and mesonic decays of various dibaryons with strangeness,
their dominant decay modes, and lifetimes are calculated.
Production estimates for BNL's Relativistic Heavy-Ion Collider are presented
employing wave function coalescence. Signals for detecting strange
dibaryon states in heavy-ion collisions and 
revealing information about the unknown hyperon-hyperon interactions are outlined.
\end{abstract}

\draft
\vspace*{-0.5cm}

Relativistic heavy-ion collisions provide a prolific source of strangeness: dozens 
of hyperons and kaons are produced in central collisions at BNL's AGS and at
CERN SPS (see e.g.\ \cite{SQM98}).
This opens the exciting perspective of forming composites with multiple units
of strangeness hitherto unachievable with conventional methods.

Exotic forms of deeply bound objects with strangeness have been proposed by
Bodmer \cite{Bodmer71} as collapsed states of matter, either consisting of
baryons or quarks.  
A six-quark bag state, the H dibaryon, was predicted by Jaffe
\cite{Jaffe77}. Other bound dibaryon states with strangeness were proposed
using quark potentials \cite{Gold87,Gold98} or the Skyrme model \cite{Schwes95}.
On the hadronic side, hypernuclei are known to exist already for a long time. 
The double $\Lambda$ hypernuclear events reported so far are closely related to
the H dibaryon \cite{Dalitz89}.
Metastable exotic multihypernuclear objects (MEMOs) 
as well as purely hyperonic systems of $\Lambda$'s and $\Xi$'s
were introduced in \cite{Scha92,Scha93} as the hadronic counterparts to
multistrange quark bags (strangelets) \cite{Gilson93,Scha97}.
Most recently, the Nijmegen soft-core potential was extended to the 
full baryon octet and bound states of $\Sigma\Sigma$, $\Sigma\Xi$, and $\Xi\Xi$ 
dibaryons were predicted \cite{Stoks99a}.

One major uncertainty for the detection of such speculative states
is their (meta)stability. 
MEMOs, for example,  consists of nucleons, $\Lambda$'s, and $\Xi$
and are metastable by virtue of Pauli-blocking effects.
Only two investigations about the weak
decay of dibaryons exist so far: 
In \cite{Don86}, the H dibaryon was found to decay dominantly by 
H$\to\Sigma^-+p$ for moderate binding energies. 
The $(\Lambda\Lambda)_b$, which has exactly the same quantum
numbers as the H dibaryon, was studied in \cite{Krivo82}. Here, the main
nonmesonic channel was found to be $(\Lambda\Lambda)_b\to\Lambda +n$.

In the following, we will revive an 'old'  approach to calculate
weak decay channels and lifetimes of various strange dibaryons using 
SU(3) symmetric contact interactions. 
Finally, we present production estimates for RHIC combining 
transport simulations using Relativistic Quantum Molecular Dynamics, 
which is widely used for simulations of
relativistic heavy ion collisions, with wave function coalescence. 

The weak decays of the octet hyperons
($\Lambda$, $\Sigma$, and $\Xi$) can be described by an effective SU(3) symmetric
interaction with a parity-violating ($A$) and a parity conserving ($B$)
amplitude \cite{Marshak}.
The weak operator is assumed to be proportional to the Gell-Mann matrix
$\lambda_6$ which ensures hypercharge violation $|\Delta Y|=1$, the $\Delta
I = 1/2$ rule and the Lee-Sugawara relation for the $A$ amplitudes.
There are three {$\cal CP$} invariant terms for the $A$ amplitude. One contributes
to $\Sigma^+\to n+\pi^+$ and can be ignored.
The two remaining parameters can be well fitted 
to the experimental data (see below).

The problem is to describe correctly the $B$ amplitudes which defy a
consistent explanation.
Traditionally, one uses the pole model which in its basic version 
is not able to describe the experimental measured amplitudes
\cite{Don86r}. Various solutions have been proposed to remedy the
situation like including the vector
meson pole \cite{Gronau72} or hyperon resonances \cite{LeYaouanc79}.
On the other hand, as pointed out in \cite{Don86r}, there is no serious
consideration about a contact interaction for  
the $B$ amplitudes in the literature. 

General SU(3) symmetry and 
{$\cal CP$} invariance results in five independent terms for the $B$ amplitudes
\cite{Marshak}. 
We find that one term gives the wrong sign either to the $B$ amplitudes for the
$\Lambda$ or for the $\Xi$'s. Hence, it must be small compared to the others.
Another term gives a contribution to $\Sigma^-\to n+\pi^-$ 
and can be neglected. 
Only three terms remain with coupling constants to be 
adjusted to the seven measured $B$ amplitudes. 

The corresponding Lagrangian for both amplitudes reads 
\begin{eqnarray}
{\cal L} &=& D {\rm Tr} \bar BB \left[P,\lambda_6\right] 
                    + F {\rm Tr} \bar B \left[P,\lambda_6\right] B
+ G {\rm Tr} \bar B P \gamma_5 B \lambda_6
\nonumber \\ &&{}
+ H {\rm Tr} \bar B \lambda_6 \gamma_5 B P 
+ J {\rm Tr} \bar B \{P,\lambda_6\} \gamma_5 B 
\end{eqnarray}
$B$ stands for the baryon octet and $P$ for the pseudoscalar nonet.
The choice $D=4.72$ and $F=-1.62$ for the $A$ amplitudes and $G=40.0$, $H=47.8$,  
and $J=-7.1$ for the $B$ amplitudes in units of $10^{-7}$ gives a good 
agreement with the experimental data as shown in Table~\ref{tab:weakamp}. 
We point out that the $B$ amplitudes do {\em 
not} follow a Lee-Sugawara relation \cite{Marshak}.
\begin{table}[tbp]
\begin{center}
\begin{tabular}{c|d|d|d|d}
 & \multicolumn{2}{c|}{A} & \multicolumn{2}{c}{B} \cr
 & exp.\ & SU(3) & exp. & SU(3) \cr
\hline
$\Lambda\to p+\pi^-$ & 3.25 & 3.25 & 22.1 & 22.1 \cr
$\Lambda\to n+\pi^0$ & $-$2.37 & $-$2.30 & $-$16.0 & $-$15.6 \cr
$\Sigma^+\to n+\pi^+$ & 0.13 & 0.0 & 42.2 & 40.0 \cr
$\Sigma^+\to p+\pi^0$ & $-$3.27 & $-$3.33 & 26.6 & 28.3 \cr
$\Sigma^-\to n+\pi^-$ & 4.27 & 4.71 & $-$1.44 & 0.0 \cr
$\Xi^0\to \Lambda+\pi^0$ & 3.43 & 3.19 & $-$12.3 & $-$11.7 \cr
$\Xi^-\to \Lambda+\pi^-$ & $-$4.51 & $-$4.51 & 16.6 & 16.6
\end{tabular}
\caption{The hyperon weak decay amplitudes in SU(3)$_{\rm weak}$ compared to
  experimental data taken from \protect\cite{Don86r}.  All values are in units
  of $10^{-7}$.}   
\label{tab:weakamp}
\end{center}
\vskip-1cm
\end{table}
Using this model for the weak hyperon decay, one can
calculate the weak mesonic and nonmesonic decay of strange dibaryons 
using a Hulthen-like wave function \cite{Krivo82}.
The meson exchange model for the weak nonmesonic decay of hypernuclei 
has been proven to be quite successful \cite{Parreno97}.
We include pion and kaon exchange in our model for the nonmesonic decay
as they are the dominant contributions. Effects from short-range contributions
like vector meson exchange \cite{Parreno97} and direct quark-quark
contributions \cite{Oka99} have been found to be less important.
We find that the p-wave
contributions originating from the $B$ amplitudes, the kaon exchange terms and
the interference terms are particularly important for the nonmesonic decay
channels. Hence, a consistent scheme of both amplitudes turns out to be a
crucial ingredience. Clearly, a more fundamental approach is desirable 
but is at present not at hand before we understand strong
interactions at the confinement scale.

\begin{figure}[tbp]
  \begin{center}
    \leavevmode
    \epsfxsize=0.46\textwidth
    \epsfbox{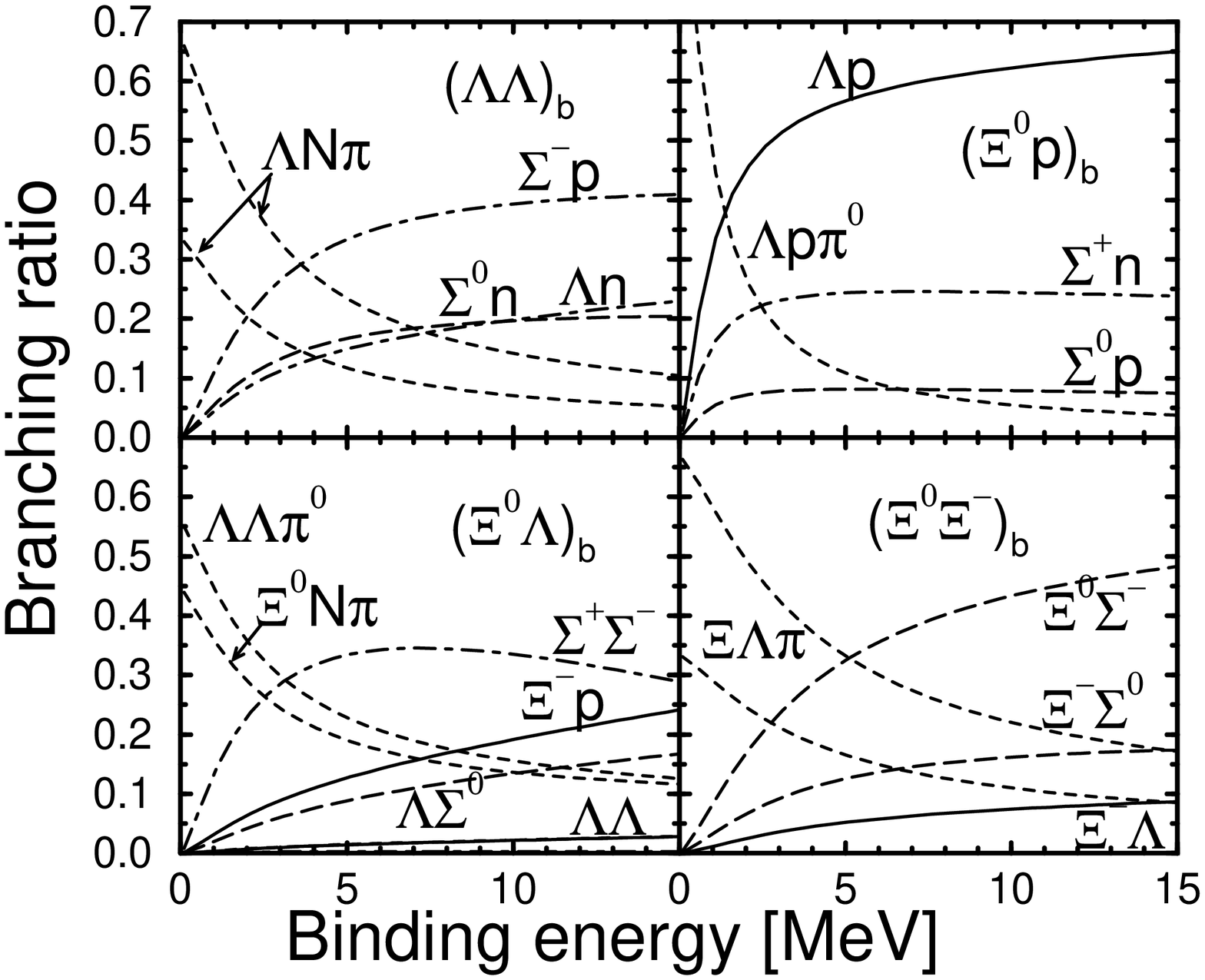}
    \vskip0.4cm
    \caption{Weak decay branching ratios for several strange dibaryons versus
      the binding energy. Solid lines denote ultimately charged final states,
      while dash-dotted lines indicate final states accessible with a neutron detector.}
    \label{fig:br_prl}
  \end{center}
\vskip-.5cm
\end{figure}

For a detection in heavy-ion experiments we are mainly interested in
candidates whose final decay products are charged:
\begin{mathletters}\begin{eqnarray}
(\Sigma^+p)_b &\to& p + p \\
(\Xi^0 p)_b &\to& p + \Lambda   \\
(\Xi^0\Lambda)_b &\to& p + \Xi^- \mbox{ or } \Lambda + \Lambda  \\
\qquad (\Xi^0\Xi^-)_b &\to& \Xi^- + \Lambda \quad .
\end{eqnarray}\end{mathletters}
We find that the decay lengths for all of the above strange dibaryons is between
$c\tau \approx 1-5$ cm.
Fig.~\ref{fig:br_prl} shows the calculated branching ratios 
as a function of the binding energy.

(a): There is only one nonmesonic decay
channel for $(\Sigma p)_b\to p+p$ which we find to be dominant above 5 MeV
binding energy. 
The dibaryon should show up in the invariant $pp$ mass spectrum 
after background subtraction from event-mixing
at $M=2.128 \mbox{ GeV }-\epsilon$ where $\epsilon$ is the binding energy.
With this method the weak decay of the lightest hypernucleus
$^3_\Lambda$H$\to^3$He$+\pi^-$ has been detected in heavy-ion collisions by 
the E864 collaboration \cite{Finch99}.

(b): For the $(\Xi^0p)_b$ bound state only one mesonic but three
different nonmesonic channels contribute.
The dominant nonmesonic decay turns out to be
$(\Xi^0 p)_b\to \Lambda +p$ already for a binding energy of 2 MeV or more.
The decay itself
resembles the one for the weak decay of the $\Xi^-$ or $\Omega^-$, which have
already been detected by several experiments (see contributions in \cite{SQM98}).
Instead of an outgoing $\pi^-$ or $K^-$ there is a proton leaving the first
weak vertex. 

(c): The dibaryon $(\Xi^0\Lambda)_b$ decays to $\Xi^-+p$ and, with a small
fraction, to two $\Lambda$'s.
Therefore, it can be seen in $\Xi^-p$ or $\Lambda\Lambda$ invariant mass plots.
One has indeed seen two-Lambda events at the AGS by experiment E896
\cite{Caines99} and experiment WA97 at the SPS has already published
two-Lambda correlation functions \cite{Jach99}. There
are plans to study the correlation of two $\Lambda$'s on an 
event-by-event basis at the STAR detector at BNL's RHIC \cite{Bellwied99}.

(d): The $(\Xi^0\Xi^-)_b$ dibaryon has been predicted to be bound \cite{Stoks99a} and
its decay to $\Xi^-+\Lambda$ has a branching ratio of a few percent. 

The other bound candidates predicted by the Nijmegen model \cite{Stoks99a} 
involve weak decays with
$\Sigma$ hyperons in the final state. If one can measure neutrons, one
is sensitive to all proposed states:
\begin{mathletters}\begin{eqnarray}
(\Sigma^-\Sigma^-)_b &\to& \Sigma^- + n + \pi^- \\  
(\Sigma^+\Sigma^+)_b &\to& \Sigma^+ + p \label{eq:sipsip}  \\  
(\Xi^0 \Sigma^+)_b &\to& \Sigma^+ + \Lambda \\
(\Xi^- \Sigma^-)_b &\to& \Sigma^- + \Sigma^- \\
(\Xi^0 \Xi^0)_b &\to& \Sigma^+ + \Xi^- \\
(\Xi^- \Xi^-)_b &\to& \Sigma^- + \Xi^- \quad .
\end{eqnarray}\end{mathletters}
In addition, one can see the nonmesonic decay involving a direct neutron in
the final state, like $(\Lambda\Lambda)_b\to \Lambda+n$
and $(\Xi^-\Lambda)_b \to \Xi^- + n$.
Thus, a possible $\Lambda n$ or $\Xi^- n$ invariant mass distribution might
reveal important information about the unknown hyperon-hyperon
interactions hitherto unaccessible by experiment.
We find that the dominant nonmesonic decay for $(\Lambda\Lambda)_b$ is 
the same as for the H dibaryon, i.e.\ $(\Lambda\Lambda)_b\to \Sigma^- +p$.
This means that the two dibaryons are indistinguishable
experimentally. 
Note, that the nonmesonic decay of the
$(\Lambda\Lambda)_b$ always involves a neutral
particle in the final state.
Searches for the H dibaryon in heavy-ion collisions are indeed sensitive for a
weak decay with a $\Sigma^-$ in the final state \cite{Hank98} and may be
utilized to look for other exotic candidates. Especially the weak decay (\ref{eq:sipsip})
looks very similar to the weak decay of the H dibaryon one is already looking for, 
but with the opposite sign for the $\Sigma$ hyperon.

\begin{figure}[tbp]
  \begin{center}
    \leavevmode
    \epsfxsize=0.4\textwidth
    \epsfbox{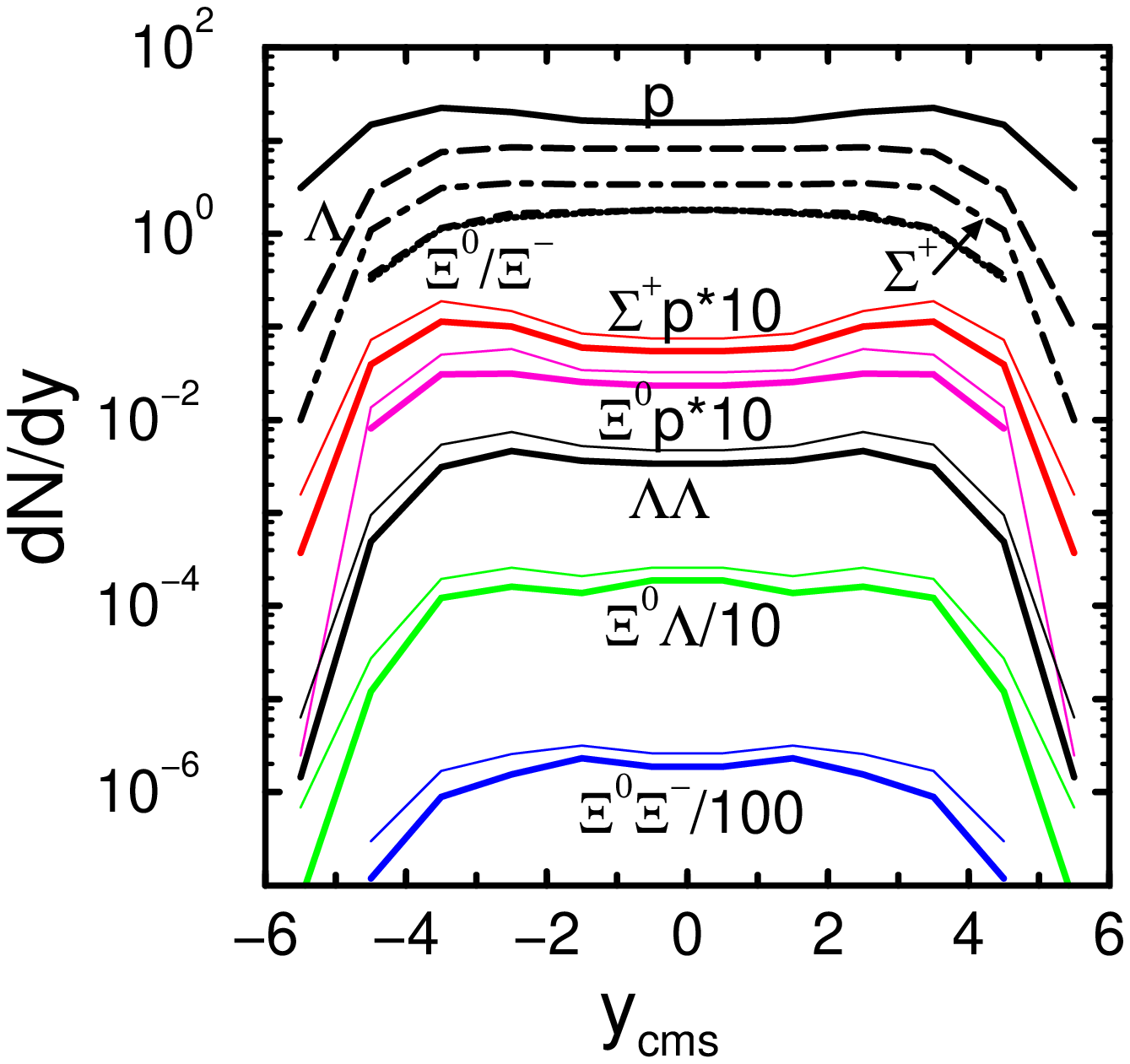}
    \caption{Rapidity distribution of baryons (upper curves) and strange
      dibaryons (lower curves) using RQMD2.4 with wavefunction coalescence for 
      Au+Au collisions at $\sqrt{s}=200$ AGeV. Upper curves are for
      a binding energy of $E_b=5$ MeV, lower ones for $E_b=1$ MeV
      ($\alpha=2$).} 
    \label{fig:coal}
  \end{center}
\vskip-.5cm
\end{figure}

Let us now focus on formation
probabilities for strange baryon clusters 
$\Lambda\Lambda$, $p\Sigma^+$, $p\Xi^0$, $\Xi^0\Lambda$ and $\Xi^0\Xi^-$.
The coalescence model provides estimates by simple
phase-space arguments. 
Momentum coalescence has been successful in describing
data at low energies (see e.g.\ \cite{Csernai86}).
At relativistic bombarding energies, however,
expansion of the source and collective flow have been shown
to strongly modify the production rates \cite{PRL,E877coal}.
Therefore, we will combine 
source distributions
for baryons borrowed from microscopic transport calculations \cite{RQMD}
with a coalescence prescription in phase space as detailed in \cite{PRC}.
This procedure has been successful to describe
deuteron yields, 
momentum distributions \cite{Jamie,PRC} and is in accord with 
studies of proton-deuteron correlations \cite{Sergei}.
Assuming uncorrelated emission the formation rate
can be expressed as
\begin{eqnarray}
{dN\over d\vec P}&=&g\int 
f_{\rm A}(\vec x_1,\vec p_1)f_{\rm B}(\vec x_2,\vec p_2)
          \rho_{\rm AB}(\Delta \vec x,\Delta \vec p) \cr
&& \delta(\vec P-\vec p_1-\vec p_2)d^3x_1d^3x_2d^3p_1d^3p_2
\end{eqnarray}
where $\Delta \vec x=\vec x_1-\vec x_2$ and $\Delta \vec p={(\vec p_1-\vec p_2)/ 2}$ 
are given in the respective
2-body cms (i.e. $\vec P\equiv0$).
One has to multiply the rate with a
symmetry factor of 1/2, if the outgoing particles are identical.
For the wave function we assume 
a Hulthen-shape as for the calculations
of weak decay properties
$\Psi (r) =  c/r \left({\rm e}^{-\kappa r} - {\rm e}^{-\alpha \kappa r} \right)$.
The statistical prefactors $g$ account for 
the lack of information about 2-body correlations with respect
to internal degrees of freedom. 
It includes spin average
and the projection on one particular final isospin state of the dibaryon.
All strange dibaryons  
are assumed to be formed in spin-singlet states.
The reduction to the correct 'number' of
possible quantum states depends crucially on the
assumption of uncorrelated emission and the nature of the bound state. 
Since the multiparticle correlations during the break up
are not well known we consider the g-values as
estimates which need further guidance -- and insight.
The predictions for strange dibaryons are depicted in Fig.\ \ref{fig:coal}.
Variations in 
the wave function parameters $E_b=1-20$ MeV ($\kappa=\sqrt{2\mu E_b}$)
and $\alpha=2-6$
lead only to minor changes in the final result ($\pm20\%$).
Therefore, we have chosen to present
calculations for the two most extreme parameter-sets
($E_b\approx 5$ MeV, $\alpha=2$) and ($E_b\approx1$ MeV, $\alpha=2$).
The formation of $\Lambda\Lambda$-states
and deuterons (see also \cite{Sorge,Monreal}) is
diminished by the volume expansion
close to midrapidity. 
For nucleon-hyperon bound states the rapidity shift
towards projectile and target is somewhat stronger
due to enhanced nuclear freeze-out densities at
forward/backward rapidities. 
Note, that strange dibaryons produced at these
rapidities have substantially longer decay lengths which opens the possibility 
of detecting them at small forward or backward angles.
 The B-parameter 
$B_{\rm AB}\propto {1\over g_{\rm AB}} N({\rm A,B})/N({\rm A})N({\rm B})$
measuring the production rates of dibaryons 
increases by a factor of two to three
comparing $\Lambda\Lambda$ and 
$\Xi\Xi$-states. 
This enhancement is compatible with an 'early' freeze-out 
scenario for multiple strange baryons as argued in \cite{Hecke98}:
Clusters with high strangeness might be formed more likely, as
they decouple earlier from the collisions zone.

There are several searches in heavy-ion collisions for the H dibaryon
\cite{Belz96,Hank98} and for long-lived
strangelets \cite{Appel96,Arm97} with high sensitivities.
Hypernuclei have been detected 
most recently in heavy-ion reactions at the AGS by the E864 collaboration
\cite{Finch99}. 
The dibaryon states studied here are short-lived. They can
in principle be detected in present and future experiments by the following
means:

1) Experiments with a time-projection chamber can track for unique exotic
decays like a charged particle decaying to two charged particles or 
tracks forming a vertex a few cm outside the target. 

2) Experiments sensitive to hyperons can look for peaks in the invariant mass
spectrum of $pp$, $p\Lambda$, $\Lambda\Lambda$, $p\Xi^-$, and $\Lambda\Xi^-$
by background subtraction using event mixing.

3) Resonances (unbound states) can be seen in the correlation function of
$\Lambda\Lambda$ \cite{GM89} and $\Lambda\Xi^-$.
Two-particle interferometry is a powerful tool 
to extract information about their (unknown) strong interaction potential as
the correlation function depends sensitively on final-state interactions 
\cite{AHR98}. 
The Coulomb potential does not mask the strong interactions
at low momenta as pointed out in \cite{Wang99} for $\Lambda p$ 
so that information about the presently unknown 
hyperon-hyperon forces can be extracted as shown in \cite{Ohnishi99}). 

The STAR experiment at the BNL's Collider RHIC is able to detect
short-lived candidates as well as exotic resonances \cite{Coffin97,Paganis99}.
One $(\Lambda\Lambda)$ resonance can be seen out of 100 uncorrelated 
$\Lambda$'s \cite{Coffin97}. For the production rates given in 
Fig.\ \ref{fig:coal} and $10^6$ central events, even the bound
$(\Xi^0\Xi^-)_b$ dibaryon can be seen by backtracking for a
reconstruction efficiency of only .2 \% or better which is indeed feasible for lifetimes around
$10^{-10}$ s \cite{Coffin97}.

In this paper we have calculated production rates of strange dibaryons via the 
coalescence mechanism of independently produced baryons.
Finally, we want to point out that another mechanism for their formation might 
be possible.  
Via the separation and 
distillation process, a hot quark-gluon plasma gets enriched with strangeness
\cite{Greiner87} leading to strangelet creation. If
strangelets are unstable they can form a doorway state by decaying to strange
dibaryons and increase the production rates.

We are pleased to acknowledge helpful discussions with
Ken Barish, Hank Crawford, Carsten Greiner, Huan
Huang, Sonja Kabana, Richard Majka, Jamie Nagle, Grazyna Odyniec, Klaus
Pretzl, Gulshan Rai, Jack Sandweiss, and Horst St\"ocker.

\end{document}